\documentclass[preprint,12pt]{elsarticle}

\usepackage{graphicx}
\usepackage{float}
\usepackage{bbold}
\usepackage{xcolor,rotating}
\usepackage{bbm}
\usepackage{dsfont}
\usepackage[colorlinks]{hyperref}
 \usepackage{hyperref}

\def\Fig#1{Fig.~\ref{#1}}

\def\0#1#2{\frac{#1}{#2}}

\hypersetup{
colorlinks=true,
citecolor=blue,
citebordercolor=red,
linktoc=all,
linkcolor=blue
}

\def\Fig#1{Fig.~\ref{#1}}

\def\0#1#2{\frac{#1}{#2}}

\usepackage{amssymb}
\journal{Nuclear Physics A}

\begin{document}

\begin{frontmatter}



\title{Towards QCD-assisted hydrodynamics for heavy-ion collision phenomenology}


\author{A. Dubla} 

\address{Physikalisches Institut,
  Universit{\"a}t Heidelberg, 69120 Heidelberg,
  Germany}
\address{GSI Helmholtzzentrum f{\"u}r
  Schwerionenforschung, 64248 Darmstadt, Germany}
  
 \author{S. Masciocchi}
 \address{Physikalisches Institut,
  Universit{\"a}t Heidelberg, 69120 Heidelberg,
  Germany}
\address{GSI Helmholtzzentrum f{\"u}r
  Schwerionenforschung, 64248 Darmstadt, Germany}
  
 \author{J. M. Pawlowski}
   \address{Institute for Theoretical Physics, Universit{\"a}t
  Heidelberg, Philosophenweg 12,  69120 Heidelberg, Germany}
  
 \author{B. Schenke}
  \address{Physics Department, Brookhaven National Laboratory,
  Upton, NY 11973, USA}
  
\author{C. Shen}
 \address{Physics Department, Brookhaven National Laboratory,
  Upton, NY 11973, USA}

\author{J. Stachel} 

\address{Physikalisches Institut,
  Universit{\"a}t Heidelberg, 69120 Heidelberg,
  Germany}

\date{\today}

\begin{abstract}
 Heavy-ion collisions are well described by a dynamical evolution
  with a long hydrodynamical phase. In this phase the properties of
  the strongly coupled quark-gluon plasma are reflected in the
  equation of state (EoS) and the transport coefficients, most
  prominently by the shear and bulk viscosity over entropy density
  ratios $\eta$/s(T) and $\zeta$/s(T), respectively.  While the EoS is
  by now known to a high accuracy, the transport coefficients and in
  particular their temperature and density dependence are not well
  known from first-principle computations yet, as well as the possible
  influence they can have once used in hydrodynamical simulations.  In
  this work, the most recent QCD-based parameters are provided as
  input to the MUSIC framework. A ratio $\eta$/s(T) computed with a QCD
  based approach is used for the first time. The
  IP-Glasma model is used to describe the initial energy density
  distribution, and UrQMD for the dilute hadronic phase. Simulations are
  performed for Pb--Pb collisions at $\sqrt{s_{\rm NN}}$ = 2.76 TeV,
  for different centrality intervals. The resulting kinematic
  distributions of the particles produced in the collisions are
  compared to data from the LHC, for several experimental
  observables. The high precision of the experimental results and the
  broad variety of observables considered allow to critically verify
  the quality of the description based on first-principle
  input to the hydrodynamic evolution.

\end{abstract}


\end{frontmatter}


%\section{}
%\label{}

In ultra-relativistic heavy-ion collisions at the Relativistic
Heavy-Ion Collider (RHIC) and at the Large Hadron Collider (LHC)
strongly-interacting matter characterised by high energy density and
temperature is produced. Under these conditions, the formation of a
deconfined state of quarks and gluons, called Quark-Gluon Plasma
(QGP), is predicted by Quantum ChromoDynamic (QCD) calculations on the
lattice at vanishing \cite{Borsanyi:2013bia,Bazavov:2014pvz} and
finite net baryon density \cite{Gunther:2016vcp,Bazavov:2017dus}.  

In the past
decade, hydrodynamic models, including viscosity, have been applied with great success to
describe the distribution of soft hadrons produced in heavy-ion
collisions at RHIC and the LHC \cite{Gale:2013da,deSouza:2015ena,Luzum:2008cw,Teaney:2009qa}.  The application
of hydrodynamics is based on the assumption of local thermal
equilibrium, energy-momentum conservation, as well as baryon number
conservation.  In this work, we will confront experimental data with a
fully integrated state-of-the-art theoretical framework.  The
IP-Glasma model is used to describe the initial energy density
distribution \cite{PhysRevLett.108.252301,Schenke:2012fw}, which provides realistic event-by-event
fluctuations and non-zero pre-equilibrium flow at the early stage of
heavy-ion collisions. Individual collision systems are evolved using
relativistic hydrodynamics with non-zero shear and bulk
viscosities. As the density of the system drops, fluid cells are
converted into hadrons and further propagated microscopically using a
hadronic cascade model (UrQMD) \cite{urqmd}.

A direct experimental study of the hydrodynamical phase of heavy-ion
collisions would reveal the QCD physics reflected in the equation of
state as well as the transport coefficients, and in particular in the
shear and bulk viscosities. However, most observables measured in
experiments are sensitive to the details and uncertainties of all the
different phases in the evolution of a heavy-ion collision. These uncertainties range
from missing knowledge about the initial state, to the details of the
kinetic phase of the dynamics and the details of the hadronisation
phase. This combination of uncertainties makes it very difficult to
reliably pin down physics in a specific phase. Still, in recent years
the large wealth of experimental data have been used to provide Bayesian estimates
for the transport coefficients \cite{Bernhard:2016tnd}.

In turn, a first-principle determination of this QCD
input would remove part of the overall uncertainties of the
description of the dynamical evolution of a heavy-ion collision. This
would give more reliable access to the physics of the other phases
such as the initial state. The QCD input for the hydrodynamic
simulations is given by the equation of state (EoS) and the transport
coefficients. For the hydrodynamical simulations in the present work
we use a state-of-the-art EoS that matches the most recent lattice EoS at
zero baryon density \cite{Borsanyi:2013bia,Bazavov:2014pvz}. This
fixes the crossover temperature to $ T_{c}\approx$ 155\,MeV,
determined from the reduced quark susceptibilities \cite{PhysRevLett.69.3282}.
These recent lattice simulations of the equation of state show a
relatively small statistical and systematical error, and the remaining
uncertainty has no significant impact on the results of the
hydrodynamical evolution. Accordingly, at zero baryon chemical potential, we consider the QCD EoS input to the hydrodynamical simulation as settled. 

The situation is very different for the transport coefficients, which
can be computed from equilibrium real-time correlation functions of
QCD. In the present work it is the shear viscosity
that follows with the Kubo formula from the spectral function of the
energy-momentum tensor \cite{doi:10.1143/JPSJ.12.570}, 
\begin{equation}
\eta=\lim_{\omega\to 0}\0{1}{20} \0{\rho_{\pi\pi}(\omega,\vec 0)}{\omega}\,.
\label{eq:Kubo}
\end{equation}
with the spectral function $\rho_{\pi\pi}$ of the spatial, traceless
part $\pi_{ij}$ of the energy-momentum tensor. As most quantitative
first-principle approaches to QCD such as the lattice or functional
methods work at imaginary time, a numerical Wick rotation to real time
is required for getting access to the transport coefficients from
these methods. Such numerical analytic continuation methods come with
large statistical and in particular systematic uncertainties. These errors
grow large in particular for low frequencies which are specifically
relevant for the computation of transport coefficients. Consequently lattice results for the shear viscosity are
only available for a few temperatures in pure Yang-Mills theory, see
e.g.\ \cite{Nakamura:2004sy,Meyer:2007ic,Mages:2015rea,Astrakhantsev:2017nrs,Pasztor:2018yae}. In
general, the computation of transport coefficients from imaginary time
data of the respective correlation functions from the lattice and
functional methods is intricate.

In the present work we utilise data for the shear viscosity over
entropy ratio from a functional diagrammatical approach to QCD
transport coefficients put forward in
\cite{Haas:2013hpa,Christiansen:2014ypa}. This approach is based on a
diagrammatic real-time representation of the correlation functions,
and circumvents the continuation problems discussed above. However, it
is subject to other systematic errors as it requires the knowledge of
real-time correlation functions of quarks and gluons, and in
particular the gluon spectral function $\rho_A$. In
\cite{Haas:2013hpa,Christiansen:2014ypa} the latter has been computed
via spectral reconstruction from imaginary time data. While at first
sight this brings back the continuation error at small frequencies,
the loop representation turns the problematic limit for
$\rho_{\pi\pi}$ in Eq. \ref{eq:Kubo} into a frequency-integral over
products of $\rho_A$. Basically it turns the multiplication with
$1/ \omega$ into a multiplication with $\omega$, hence suppressing the
low frequency regime of the reconstructed $\rho_A$ in the computation
of $\eta$. 

Still the approach has its own systematic uncertainties: the dominating
one results from the
approximate computation of the real-time correlation functions of
quarks and gluons, and in particular the well-known problems at higher
temperatures. The reconstruction of the relevant thermal part is
performed only on the first few Matsubara frequencies. Typically this
leads to an unphysical broadening of the spectral function at large
temperatures. In the present case this would lead to a smaller thermal
slope of $\eta$/s(T) at large temperatures $T/T_c \geq 1$.

A second relevant source for the systematic uncertainty is the neglection of
higher order diagrams. After a resummation, the diagrammatical
representation consists of one- to three-loop diagrams of full
correlation functions. Note that the loops here have nothing to do
with a perturbative ordering. In
\cite{Haas:2013hpa,Christiansen:2014ypa} it has been argued that the
higher order diagrams are subject to a very efficient phase space
suppression that originates in the quasi-particle structure of the
gluon spectral function. In \cite{Christiansen:2014ypa} all one- and
two-loop diagrams have been computed. The rapid convergence argument
has been checked by computing the dominant three-loop diagrams, the
total value of which is in the permille regime. However, while the
resummation works very efficiently it also leads to a global
normalisation problem also discussed in \cite{Fister:2013bh}. It is
related to a standard multiplicative renormalisation. It gives rise to
the biggest systematic error in this approach, and will be treated in
a future publication.

\begin{figure}[!ht]
\centering
\includegraphics[scale=0.6]{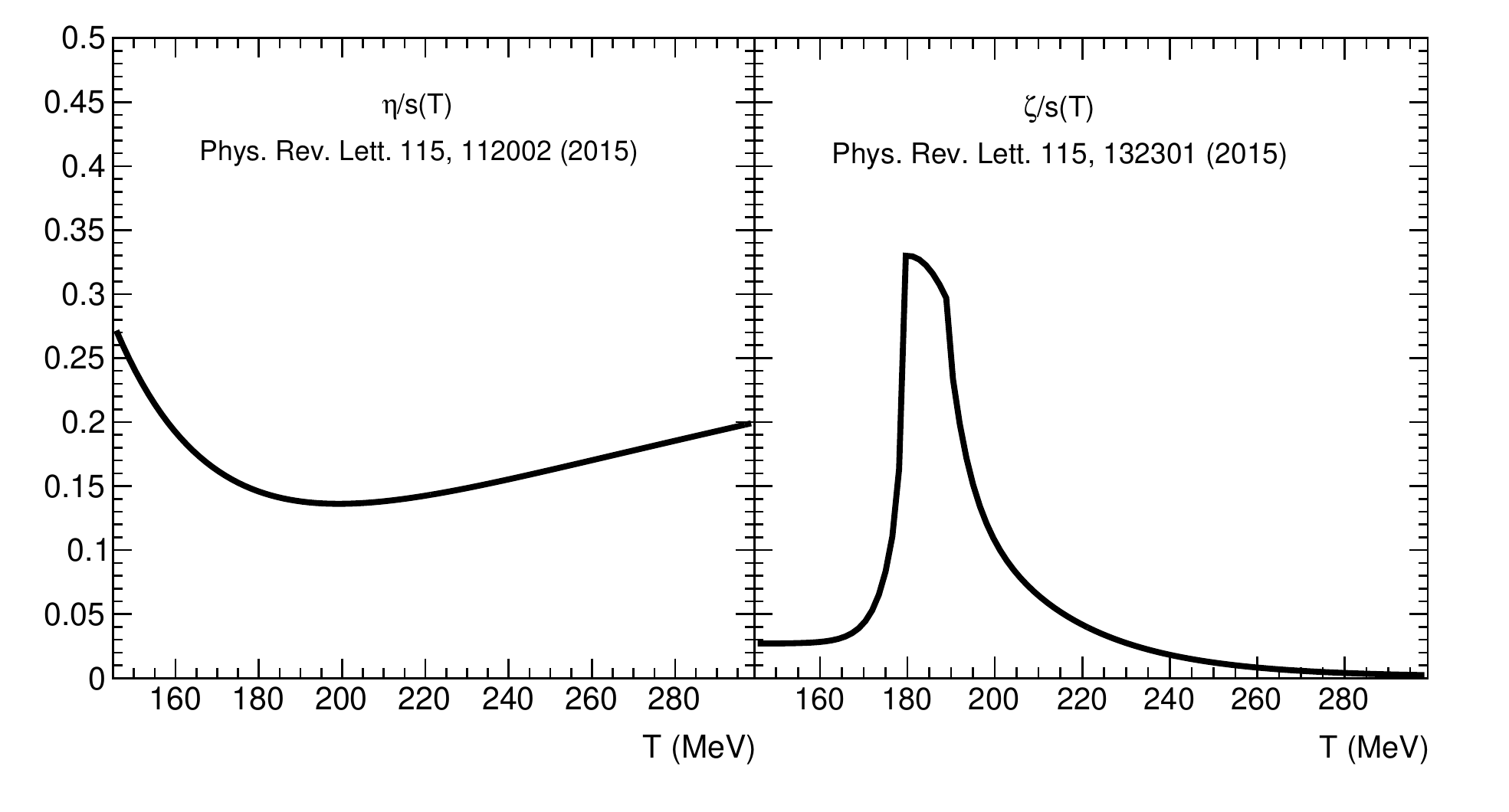}
\caption{Left panel: Analytic function for $\eta$/s(T) as a
  function of temperature as given in Eq. \ref{eq:etas}. Right panel:
  The $\zeta$/s(T) parametrization as a function of temperature.}
\label{trcoef}
\end{figure}

In the present work we use pure glue $\eta$/s(T) as computed
\cite{Haas:2013hpa,Christiansen:2014ypa}. We estimate the impact of
the combined systematic error by $\eta/s(T)\to \eta/s(T) +d$ with a
temperature-independent shift $d$. The result in
\cite{Christiansen:2014ypa} is parameterised well with
%\begin{equation}  
%\eta/s(T) = \frac{a}{\alpha_{s,{\text{\tiny{HQ}}}}^{\gamma}(c\,T /T_{c})} + \frac{b}{(T/T_c)^{\delta}},
 %   \label{eq:etas}
%\end{equation}
\begin{equation}  
\eta/s(T) = \frac{a}{\alpha_{s,{{\rm \tiny{HQ}}}}^{\gamma}(c\,T /T_{c})} + \frac{b}{(T/T_c)^{\delta}},
    \label{eq:etas}
\end{equation}
where a = 0.15, b = 0.14, c = 0.66 and the scaling coefficients
$\gamma=1.6$ and $\delta$ = 5.1. The strong
coupling in Eq. \ref{eq:etas} is a heavy quark effective coupling with a
simple analytic form \cite{Nesterenko:1999np}
\begin{equation}
\alpha_{s,{{\rm \tiny{HQ}}}}(z)= \0{1}{\beta_0} \0{z^2-1} {z^2 \log z^2} 
\label{eq:HQ-coupling}\end{equation} 
We emphasise that the coupling present in the numerical computation in
\cite{Haas:2013hpa,Christiansen:2014ypa} is that of the underlying
functional approach to QCD at finite temperature (see
\cite{Fister:2013bh,Cyrol:2017qkl}). The above fit of $\eta$/s(T) is
depicted in the left panel of \Fig{trcoef}. It shows a minimum at
$\rm T_{min}~\approx$ 1.26 $\rm T_{c}$ with a value of 0.14.  In our
simulations we varied this value with the shift $d\in[-0.06, 0]$
between 0.14 and the AdS value 0.08 for an estimate of the impact of
the systematic error on $\eta/s$. Moreover, in
\cite{Christiansen:2014ypa} a QCD estimate has been provided with a =
0.2, b = 0.15, c = 0.79. As the effect of changing $\eta$/s(T) from
the pure glue result to the QCD estimate is covered by the uncertainty
estimate obtained from varying the shift parameter $d$ we refrain from
discussing the related results separately.  We have also checked that
a multiplication with a temperature-independent factor has an effect
similar to such a shift.

It is remarkable that the fit Eq. \ref{eq:etas} works so well as it is a
direct sum of a hadron resonance gas-type behaviour for
$T\lesssim T_c$ and a hard-thermal-loop--type behaviour for
$T\gtrsim T_c$. Note however that in the latter non-perturbative
information is carried by the running coupling. In any case this
indicates a rapid transition in the hydrodynamical evolution from the
hadronic phase to the quark-gluon phase. This behaviour is also seen
for the gluonic Debye mass \cite{Cyrol:2017qkl} in the functional
approach underlying the computations in \cite{Christiansen:2014ypa} as
well as further thermal observables. If this rapid transition is also
seen for other transport coefficients it would support the common
use of simple fits for transport coefficients.  The computation and
use of other transport coefficients is left to future work. A
respective computation of the bulk viscosity is under way, for recent lattice results 
see \cite{Astrakhantsev:2017nrs}.
Due to the
diagrammatic similarity of the computation, the systematic uncertainties are
correlated. Hence, the multiplicative renormalisation factor drops out
in the ratio $\zeta/\eta(T)$.

In a previous publication \cite{bulk} it has been observed that the
bulk viscosity $\zeta$/s(T) is important for describing the
experimentally observed radial flow and azimuthal anisotropy
simultaneously.  In the absence of first-principle results for
$\zeta$/s(T) we take over the same functional form of $\zeta$/s(T)
as in \cite{bulk2}. It is depicted in the right panel of
Fig. \ref{trcoef}.  The bulk viscosity coefficient shows a maximum at
T = 180 MeV and starts to decrease almost exponentially as the system
cools down.  At the maximum, the value of $\zeta$/s(T) is $\approx$
0.3. The coefficient vanishes in the high temperature limit in the QGP
phase whereas in the low temperature limit (hadronic gas), it
converges to a finite value equal to 0.03.

In the switching between hydrodynamics and hadron cascade simulations
the iSS particle sampler \cite{iss2,iss1} converts the hydrodynamic outputs on the
hyper-surface into various hadrons (particlization) with specific
momenta and positions. Particlization denotes the conversion of the
hadronic medium from macroscopic to microscopic degrees of
freedom. In this work the $T_{switch}$ between the hydro and
UrQMD phases is set to $T_{switch}$ = 145 MeV.  More specifically,
such a Monte Carlo event generator is constructed according to the
differential Cooper-Frye formula:

\begin{equation}
  E \frac{dN_{i}}{d^{3}p} = \frac{gi}{(2\pi)^{3}} \int_{\Sigma}^{\, } f_{i}(x, p) p^{\mu} d^{3}\sigma_{\mu},
\end{equation}

\noindent where $f_{i}$ is the distribution function of particle $i$, which
includes both equilibrium and non-equilibrium contributions
$f_{i}\, =\, f_{0,i} + \delta f_{i}$, integrated in a volume element of
the switching hypersurface $\Sigma$, defined by the
constant switching temperature $T_{switch}$.  The non-equilibrium
distribution function is composed by two terms $\delta f$ =
$\delta f_{shear}$ + $\delta
f_{bulk}$. 
The same functional forms for 
$\delta f_{shear}$ and $\delta
f_{bulk}$ as in \cite{bulk} have been used,

\begin{equation}
\delta f_{shear} = f_{0}(1 \pm f_{0})\frac{\pi^{\mu \nu}p_{\mu}p_{\nu}}{2T^2(e + P)} 
\end{equation}

and

\begin{equation}
\delta f_{bulk} = f_{0}(1 \pm f_{0})\frac{-\Pi}{\zeta/\tau_{\Pi}} \frac{1}{3T} \bigg(\frac{m^2}{E} - (1 - 3c_s^2)E\bigg), 
\end{equation}

\noindent where $\pi^{\mu \nu}$ is the shear stress tensor, $\Pi$ is the bulk pressure,
and $\tau_{\Pi}$ is the relaxation time for bulk viscosity.

The bulk viscous correction influences the
particle spectra and the flow observables when the expansion rate is large. We want to emphasize that
the non-equilibrium corrections of the bulk viscosity to the momentum
distribution of hadrons at the moment of switching are still not
completely understood from a theoretical point of view and represent a
large source of uncertainty in the simulation. Specific effects on measured observables from the bulk
viscous correction have been discussed in \cite{bulkcorrec}.

In this paper, we study multiplicities of different
hadron species, transverse momentum ($p_{\rm T}$) differential spectra and various flow
observables in Pb--Pb collisions at 2.76 TeV for the centrality
intervals 0--5\%, 5--10\%, 10--20\%, 20--30\% and 30--40\%.

\begin{figure*}[ht!]
\centering
\includegraphics[scale=0.7]{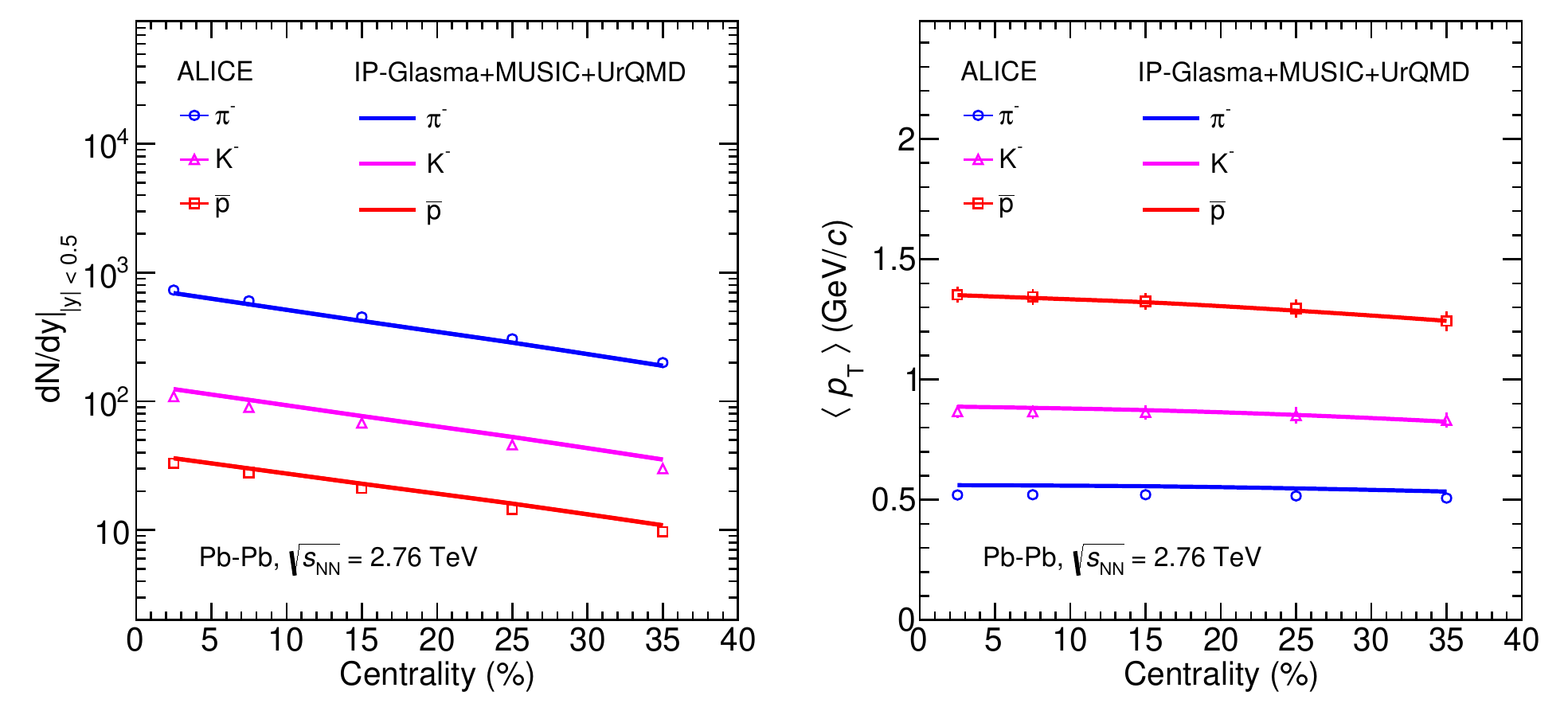}

\centering
\includegraphics[scale=0.35]{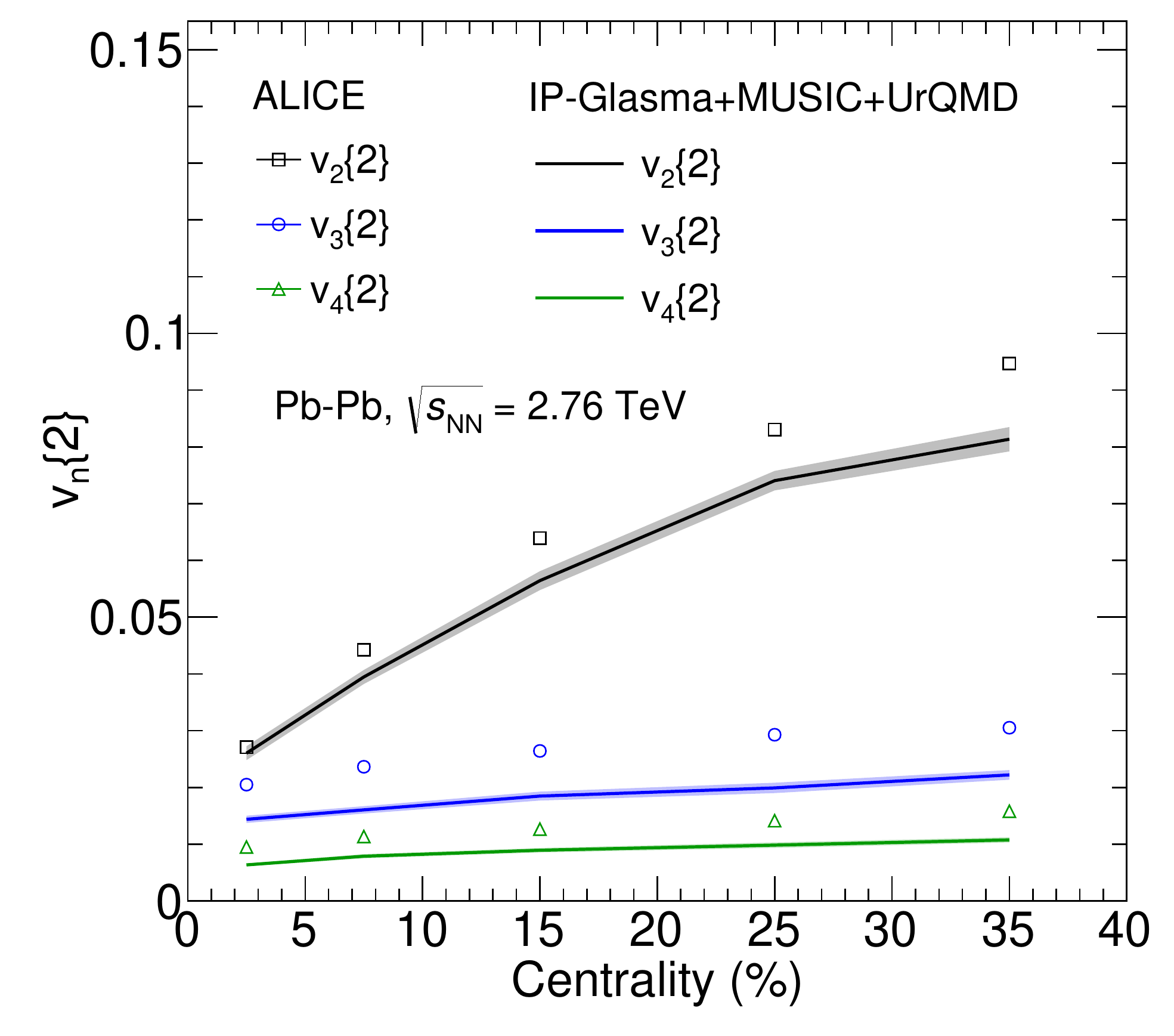}
 \caption{ Mid-rapidity densities dN/d$y|y<$0.5 (top left panel), mean--$p_{\rm T}$ (top right panel) and $p_{\rm T}$-integrated $v_{n}\{2\}$ (bottom panel) as functions of centrality from the currect calculation in comparison with the ALICE experimental measurements~\cite{alicemeanpt}.}
\label{centobs}
\end{figure*}

The IP-Glasma model give a good description of the centrality dependence of the
charged hadron multiplicity \cite{Schenke:2012fw}. 
It is observed that charged hadron multiplicities at mid-rapidity as a function
of collision centrality for Pb--Pb collisions at 2.76 TeV obtained from the simulations are in good
agreement with the experimental measurements performed by ALICE in
Pb--Pb collisions at 2.76 TeV \cite{alicemult}. In the
left top panel of Fig. \ref{centobs} we also compare our results of identified particle yields at mid-rapidity as a function
of collision centrality  for
pions, kaons, and protons with the ALICE
measurements~\cite{alicemeanpt}. Our calculation agrees with the
experimental data within the uncertainties and we observed that
also the centrality dependence is well reproduced for identified hadrons.

In the right top panel of Fig. \ref{centobs}, the comparison of the
mean $p_{\rm T}$, $\langle p_{\rm T} \rangle$, for pions, kaons, and
protons as a function of centrality from our simulations is compared
with the ALICE experimental measurements~\cite{alicemeanpt}. As it is
explained in \cite{bulk, hadrresc}, the bulk viscosity and the UrQMD
phase play a fundamental role in the simulation. Without them
we would not be able to have a good description of the
$\langle p_{\rm T} \rangle$ and of the $p_{\rm T}$-differential hadron
spectra discussed later.  The change in $\langle p_{\rm T} \rangle$
due to the inclusion of the bulk viscosity is significant for all
hadron species (pions, kaons, and protons) \cite{bulk}. Bulk viscosity is
essential for a simultaneous description of the multiplicity and
$\langle p_{\rm T} \rangle$ of hadrons when IP-Glasma initial
conditions are used. Without the bulk viscosity, the system expands
too rapidly, leading to a too large hydrodynamic transverse flow. Bulk
viscosity improves the agreement with data by acting as a resistance
to expansion, reducing the transverse flow of the system.  The
$\langle p_{\rm T} \rangle$ of protons agrees with experimental
measurements through all centralities, and the values are observed to
increase with respect to the results obtained from the simulations in which UrQMD was not used. This reflects that
the more massive protons show a large sensitivity of hydrodynamic radial flow
as it was already observed in \cite{hadrresc, vishnu}. The effect of
hadronic rescattering is very similar across centrality classes.
Although the pion $\langle p_{\rm T} \rangle$ is barely affected by
the hadronic rescatterings, it is observed that the
$\langle p_{\rm T} \rangle$ in more central collisions is slightly
overestimated in the simulations.

It is important to mention that a rapid change of the transport coefficients likely happens when switching to UrQMD. In UrQMD \cite{Demir:2008tr} (as well as in the most recent SMASH transport code \cite{Rose:2017bjz}), ($\eta$/s)($T_{\rm switch}$) $\approx$ 1, while ($\eta/$s)($T_{\rm switch}$) used in our simulation has a value of ~0.27, as shown in Fig. \ref{trcoef}. However, charged hadron observables, as well as identified pions and kaons are very insensitive to the details of the microscopic hadronic description (see e.g. \cite{hadrresc}), and only protons are potentially affected by this discontinuity in the transport coefficients and uncertainties in the microscopic hadronic transport simulation.

In the bottom central panel of Fig. \ref{centobs} the $p_{\rm T}$-integrated charged hadron
anisotropic flow coefficients, $v_{2,3,4}\{2\}$, in Pb--Pb
collisions at 2.76 TeV are shown as a function of centrality and are compared with the ALICE measurements~\cite{PhysRevLett.107.032301}. 
We calculate the flow harmonics $v_n\{2\}$ using the
2-particle cumulant method within 0.2 $< $ $p_{\rm T}$ $<$ 5.0 GeV and
$|\eta|$ $<$ 0.8, together with a pseudo rapidity gap $|\Delta \eta|$
$>$ 1.0. The band width in the simulation results represents the statistical uncertainty. It is immediately visible that the $v_n\{2\}$ coefficients are
too low in the simulation, especially when moving to more peripheral collision centrality classes. 
This is because $\eta$/s(T) is likely too large in the region below T = 200 MeV.
Temperature dependent $\eta$/s(T) values have been studied previously \cite{Bernhard:2016tnd, Niemi:2015qia, Niemi:2011ix, Denicol:2015nhu}. Besides the difference in the detailed shape, our input values differ from simple parametrizations mainly in the location of the minimum.

The third and fourth flow harmonic coefficients are
observed to be even more suppressed with respect to the second one, also in the most central collision events. 
This is expected because higher harmonics are more sensitive to the shear viscosity and they get
reduced more easily if a too large shear viscosity over entropy ratio is used in the simulations \cite{higerflowharm}.

\begin{figure*}[ht!]
\centering
\includegraphics[scale=0.65]{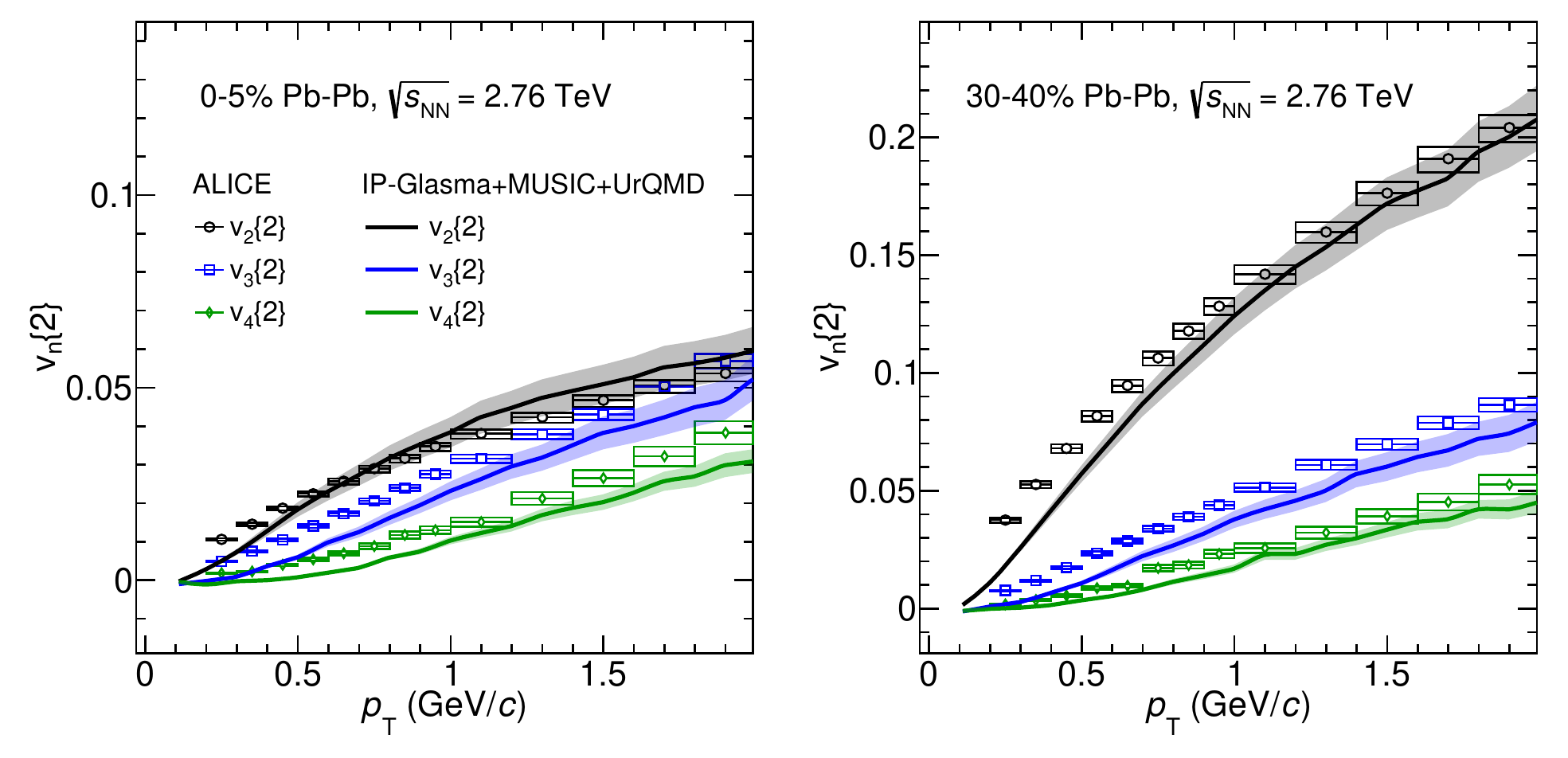}
\caption{$v_{n}\{2\}(p_{\rm T}$) (n=2,3,4) for charged hadrons in the 0--5\% (left) and 30--40\% (right) centrality intervals in comparison with ALICE measurements \cite{Abelev:2012di}.}
\label{vn}
\end{figure*}

Figure \ref{vn} shows the $p_{\rm T}$-differential flow harmonics
$v_{n}\{2\}(p_{\rm T}$) (n=2,3,4) of charged hadrons in 0--5\% (left)
and 30--40\% (right) Pb--Pb collisions at 2.76 TeV computed in our
simulation and compared with the experimental measurements by
ALICE \cite{Abelev:2012di}.  In agreement with what is discussed for
the $p_{\rm T}$ integrated $v_{n}\{2\}$ measurement, we observe that
the results of the simulation tend to underestimate the
$v_{n}\{2\}(p_{\rm T}$) coefficients in the low transverse momentum
region, where the majority of particles is produced, while a better
description of the experimental measurements is observed for
$p_{\rm T}$ larger than 1 GeV/$c$. Also a better description of the
measurements is observed in the 0--5\% most central collisions with
respect to the more peripheral case.  

\begin{figure*}[ht!]
\centering
\includegraphics[scale=0.6]{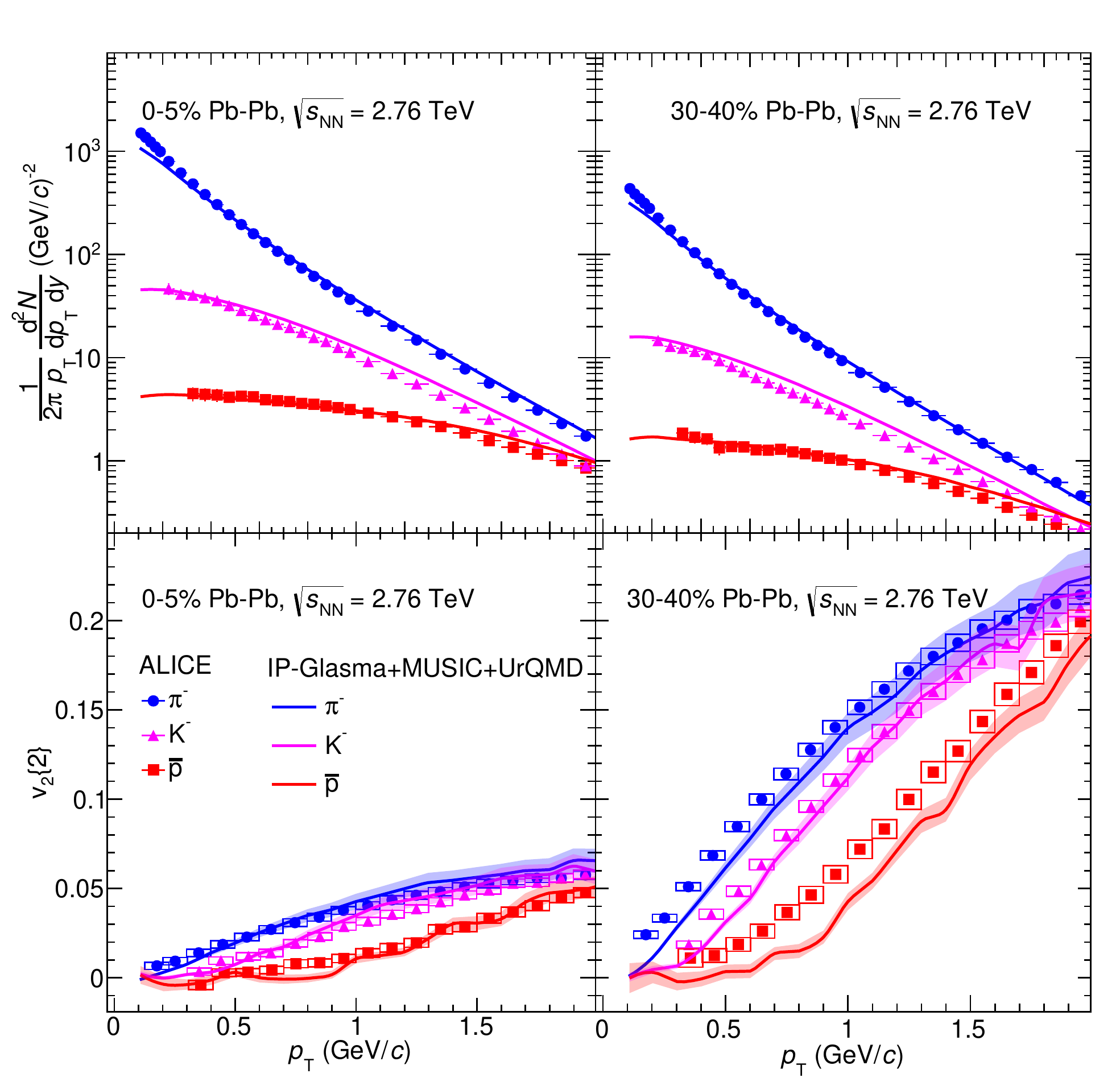}
\caption{$p_{\rm T}$ spectra (upper panels) and $v_{2}\{2\}(p_{\rm T})$ 
(lower panels) as a function of the transverse momentum of pions, kaons, and protons in comparison with ALICE measurements ~\cite{alicemeanpt, Abelev:2014pua}.  Two centrality classes are considered:  0--5\% (left) and 30--40\% (right).}
\label{ptspectra}
\end{figure*}

Beyond studying and predicting
the flow observables for charged hadrons, it is important to check the
$p_{\rm T}$ spectra of identified hadrons
since their mass dependence reflects the radial flow of the expanding
system. Fig. \ref{ptspectra} shows the results for pions, kaons, and
protons as a function of $p_{\rm T}$. In the top left panel the
invariant yields for the 0--5\% most central Pb--Pb collisions are
shown, while the yields for the 30--40\% centrality interval are shown
in the top right plot.  The calculations, including the hadronic
rescattering, agree reasonably well with the measurements. Tension with
data appears, and increases slightly in more peripheral collisions,
especially for kaons and for low-$p_{\rm T}$ pions. The better
agreement of the protons with respect to the other hadron species is
an additional indication that heavier particles show a large
sensitivity to the additional radial flow introduced by the hadronic rescattering and $\rm{B\bar{B}}$ annihilation in the UrQMD
phase; the low-$p_{\rm T}$ region of the spectra is reduced in the
transport phase as a results of shifting more protons to
higher $p_{\rm T}$.  The left and right bottom panels of
Fig. \ref{ptspectra} show the differential flow harmonics
$v_{2}\{2\}(p_{\rm T}$) for pions, kaons, and protons in 0--5\% and
30--40\% Pb--Pb collisions at 2.76 TeV compared with measurements by ALICE
using the 2-particle cumulant method within $|\eta|$ $<$ 0.8
\cite{Abelev:2014pua}. The interplay of radial and elliptic flow is
expected to lead to a dependence of the $p_{\rm T}$-differential flow
on the mass of the particle species. A clear mass ordering is observed
when comparing $v_{2}\{2\}(p_{\rm T}$) among different particle
species. We observe that the
hadronic rescattering has a large effect on the proton
$v_{2}\{2\}(p_{\rm T}$). In the 30--40\% centrality the proton
$v_{2}\{2\}(p_{\rm T}$) is pushed to higher $p_{\rm T}$ values with
respect to what is observed in data. The balance between the radial
and elliptic flow seems to be better described in the most central
collisions.  
Even though the elliptic flow around the mean--$p_{\rm T}$
is well reproduced, our calculations underestimate the
$v_{2}\{2\}(p_{\rm T}$) of pions and kaons at low $p_{\rm T}$ for more peripheral collisions. The under estimation of the $v_{2}\{2\}(p_{\rm T}$) is due to the important role played by the out-of-equilibrium correction
from bulk viscosity and also 
due to the possibly too high shear viscosity over entropy ratio, which
suppress the flow coefficients. 

\begin{figure*}[ht!]
\centering
\includegraphics[scale=0.6]{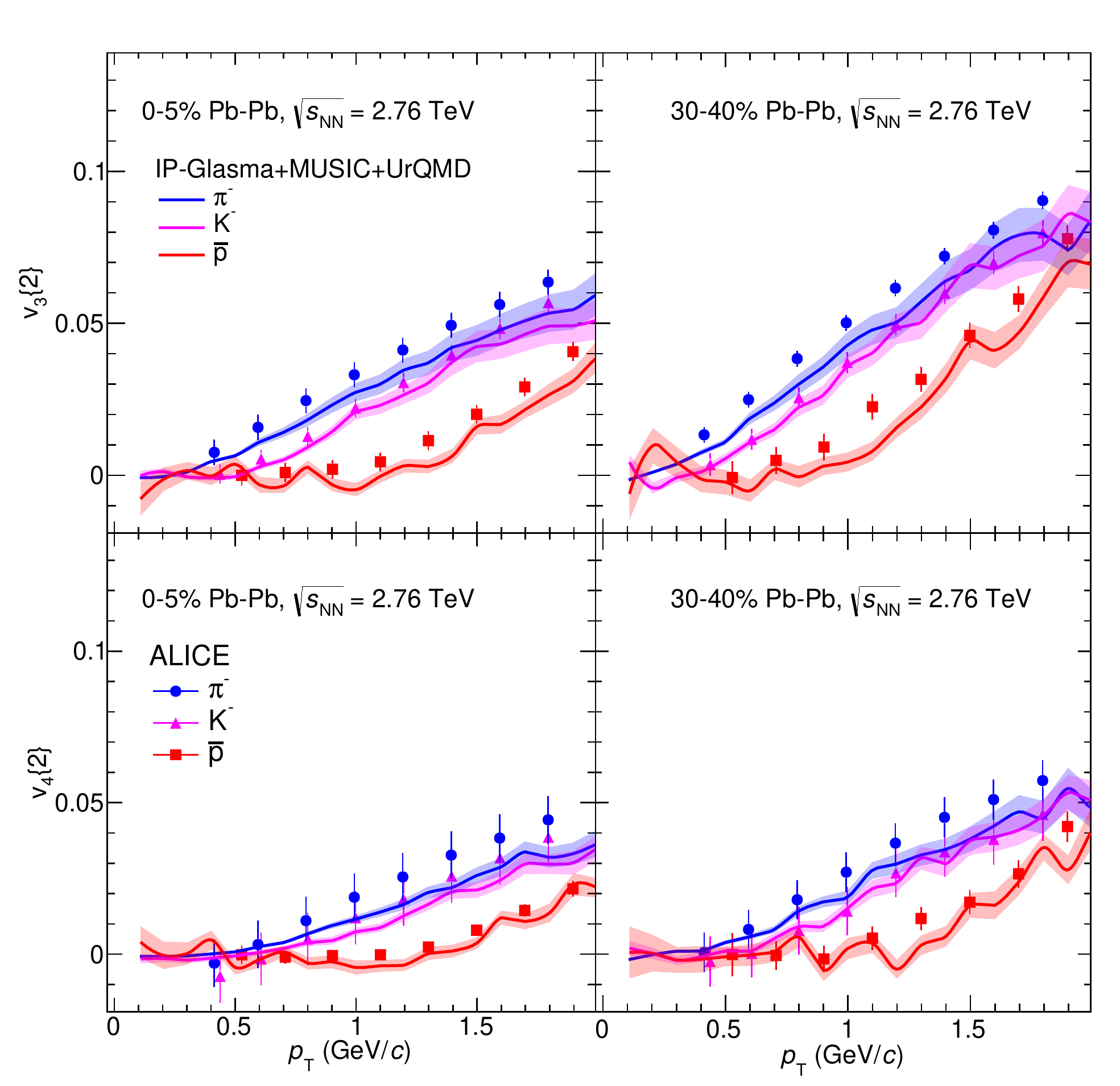}
\caption{$v_{3}\{2\}(p_{\rm T})$ (upper panels) and $v_{4}\{2\}(p_{\rm T})$ (lower panels) as a function of the transverse momentum of pions, kaons, and protons in comparison with the ALICE measurements \cite{Adam:2016nfo}.  Two centrality classes are considered:  0--5\% (left) and 30--40\% (right).}
\label{vnpid}
\end{figure*}

The upper (lower) panels of Fig. \ref{vnpid} show our calculations for
the identified hadron $v_{3}\{2\}$ and $v_{4}\{2\}$ in the two
centrality classes compared with the experimental measurements by
ALICE \cite{Adam:2016nfo}.  Note that in the ALICE data the residual
non-flow contributions have been subtracted using information from pp collisions. This is not necessary for our
calculations since the related non-flow effects are mainly from
resonance decays.  We observe that the level of agreement with data
is slightly worse for $v_{3}\{2\}$ and $v_{4}\{2\}$ compared to what is
observed for $v_{2}\{2\}$. This is because the higher harmonics, as
previously explained, are more sensitive to the shear viscosity over
entropy density ratio, and they get substantially reduced in case this transport
coefficient implemented in the simulation gets too high. The overall effect of the hadronic rescattering is similar
for $v_{3}\{2\}$ and $v_{4}\{2\}$ compared to $v_{2}\{2\}$.  As
expected, the usual mass
ordering is observed also for $v_{3}\{2\}$ and $v_{4}\{2\}$.

Let us now come back to the systematic error on $\eta$/s(T). As
discussed in Eq. \ref{eq:etas} the largest uncertainties in
the theoretical computations in
\cite{Haas:2013hpa,Christiansen:2014ypa} mainly affect the absolute
normalization of $\eta$/s(T). In order to test the effect of the absolute normalisation of the $\eta$/s(T), we shift the whole distribution according to
\begin{equation}
\eta/s(T) \to \eta/s(T)  +d\,,\qquad d\in[-0.06, 0]\,. 
\end{equation}
This covers the conjecture of minimal value of 1/4$\pi$ = 0.08, the AdS/CFT bound
\cite{Kovtun:2004de}. Note however that the AdS/CFT value is only
obtained by reducing the value of $\eta$/s(T) at the minimum by a
factor 1.75, which is a stretch of the theoretical systematic uncertainty. 

\begin{figure*}[ht!]
\centering
\includegraphics[scale=0.5]{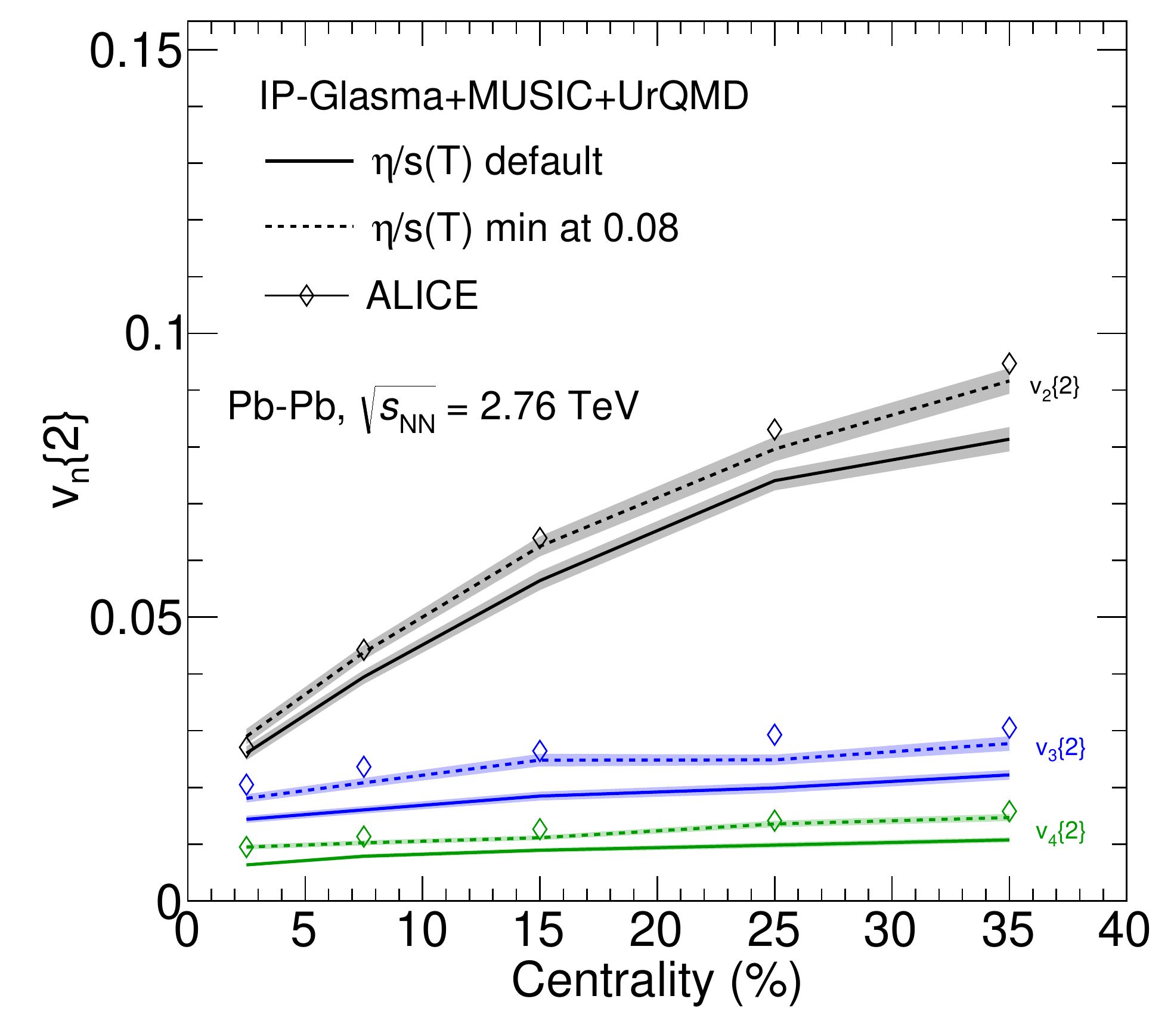}
 \caption{ $v_{n}\{2\}$ coefficients as functions of centrality measured with the default and shifted $\eta$/s(T). }
\label{centobs2}
\end{figure*}

We first present the effect of a reduced $\eta$/s(T) on the $p_{\rm T}$-integrated
charged hadron anisotropic flow coefficients $v_{2,3,4}\{2\}$ as a
function of centrality, since those are the observables that mainly
showed a disagreement with the experimental data, pointing to a too strong
$\eta$/s(T).  In Fig. \ref{centobs2} the results coming from the
simulations, using both the standard $\eta$/s(T) (solid line) and the
shifted one (dashed line), in which the shift $d$ was choosen to be 0.056, are compared with the experimental
measurements. It is observed that with the modified $\eta$/s(T) the
results of our calculations are in agreement, within the statistical
uncertainties, with the
experimental measurements by ALICE~\cite{PhysRevLett.107.032301}.
The multiplicity, the $\langle p_{\rm T} \rangle$ as well as the $p_{\rm T}$-differential spectra are not shown here because it is observed that those observables do not change in the simulations, showing less sensitivity to the $\eta$/s(T) with respect to the $v_{n}\{2\}$ coefficients.

\begin{figure*}[ht!]
\centering
\includegraphics[scale=0.65]{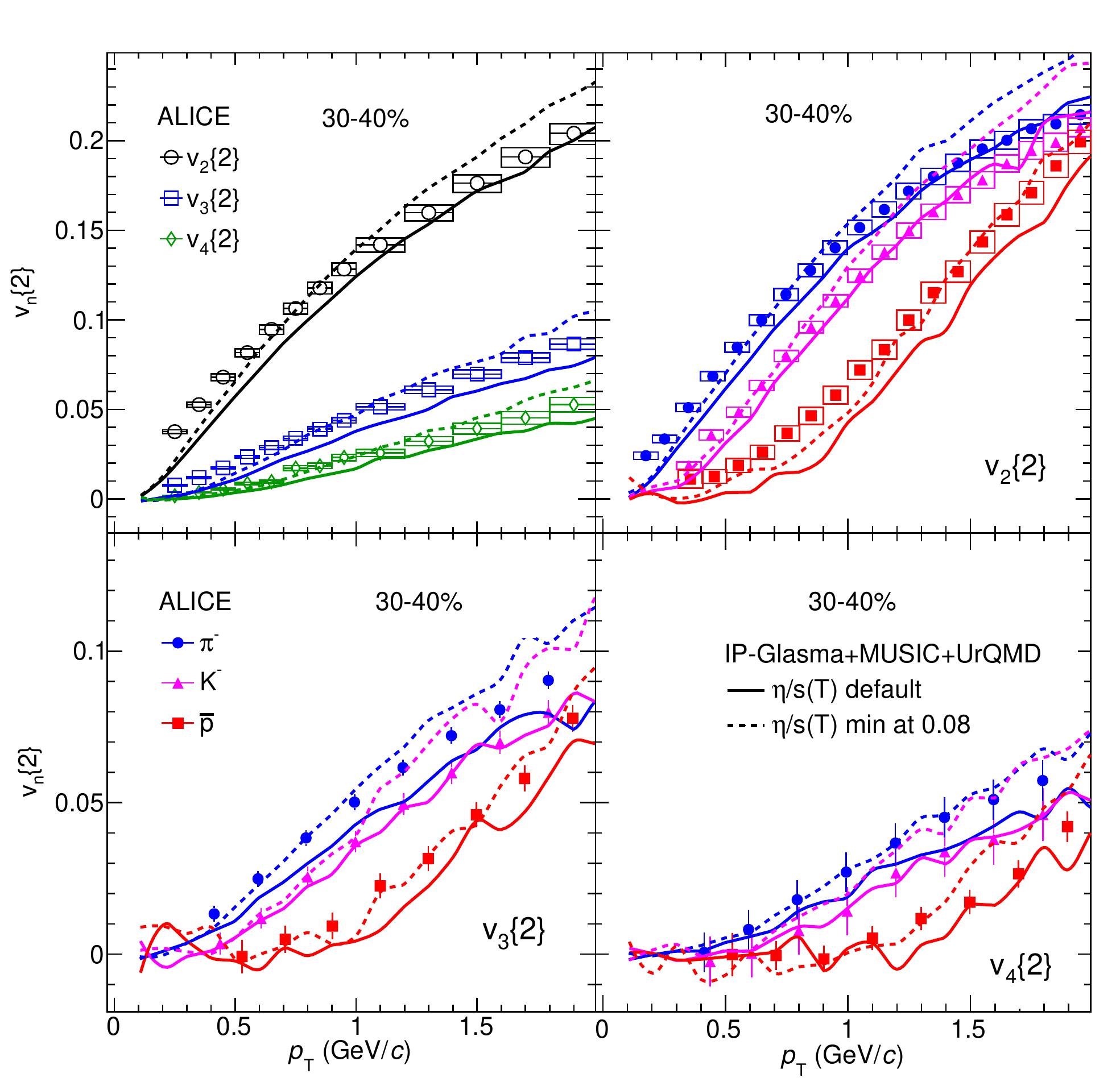}
\caption{$v_{2}\{n\}(p_{\rm T})$ (upper left panel) for charged hadrons and $v_{2}\{2\}(p_{\rm T})$ (upper right panel), $v_{3}\{2\}(p_{\rm T})$ (lower left panel) and $v_{4}\{2\}(p_{\rm T})$ (lower right panel) of pions, kaons, and protons.  The 30--40\% centrality interval is shown for all panels. Results from both parametrisation of $\eta$/s(T) are reported.}
\label{ptdiffetas2}
\end{figure*}

However, it is worth while to compare also the $p_{\rm T}$-differential flow harmonic coefficients for charged hadrons and for pions, kaons, and protons.
In the top left panel of Fig. \ref{ptdiffetas2} our calculations, using both the standard $\eta$/s(T) (solid lines) and the modified one (dashed lines), are compared with the ALICE measurements for $v_{2,3,4}\{2\}(p_{\rm T})$. Statistical uncertainties in the model calculations are not shown in Fig. \ref{ptdiffetas2}, but they can be judge from the bin-by-bin fluctuations. 
Although the flow coefficients obtained with the modified $\eta$/s(T) deviate from the data at high $p_{\rm T}$, those results show a better agreement with data for $p_{\rm T}$ $<$ 1 GeV/$c$, where most particles are produced. 
We next turn to identified hadron flow coefficients and also here we note that tensions with ALICE measurements are observed at high $p_{\rm T}$ for the $v_{2,3,4}\{2\}(p_{\rm T})$ of pions, kaons, and protons as shown in the right top panel Fig. \ref{ptdiffetas2} for $v_{2}\{2\}$, bottom left panel for $v_{3}\{2\}$ and bottom right panel for $v_{4}\{2\}$. We note that tension with measurements at high $p_{\rm T}$ is less worrisome than in lower regions of transverse momenta, since the high $p_{\rm T}$ region is more sensitive to uncertainties in the viscous corrections to the hadron distribution function ($\delta f$ of the shear and of the bulk). We note that the bulk viscosity significantly affects the entire $p_{\rm T}$ range of $v_n( p_{\rm T})$, mainly via the out-of-equilibrium correction to the distribution function $\delta f$.

In this paper, we compare results from a hybrid model of IP-Glasma initial conditions, shear and bulk viscous hydrodynamics (MUSIC), and microscopic hadronic transport (UrQMD) with a wide range of integrated and differential measurements in Pb--Pb collisions at 2.76 TeV. For the first time the shear viscosity over entropy density ratio as a function of temperature from a functional diagrammatical approach to QCD transport coefficients has been used in a state-of-the-art hydrodynamical framework.

Considering the much larger required statistics in the model calculation, we leave the computation of the differential flow harmonics $v_{n}\{2\}(p_{\rm T}$)
of $\Lambda$, $\Xi$, $\Omega$ and $\phi$ \cite{vishnu, hadrresc, bulkcorrec}, to future work.
It will be particularly interesting to study the effects of the microscopic hadronic simulation on strange and multi-strange particles. 
In the hadronic
cascade model a small hadronic cross section is assigned to strange
hadrons, and in UrQMD the magnitude of the hadronic
re-interaction is related to the number of strange quarks contained in
the hadron.
Hence the amount of radial flow that the hadrons pick up in the hadronic phase depends on their strange quark content. 
In addition, the prediction for the $\phi$-meson (containing an $s$ and $\bar{s}$ quark) was observed to be almost not affected by the hadronic phase in the UrQMD, and this was leading to the breaking of the mass ordering when comparing  $v_{2}(p_{\rm T}$) of the $\phi$-meson and proton \cite{bulkcorrec}. Due to its small hadronic cross section in UrQMD the $\phi$-meson is rather weakly coupled to the hadronic medium and it decouples from the system almost immediately after hadronization.

On the theory side, a threefold way to QCD transport coefficients is
currently pursued. Firstly, the non-perturbative diagrammatic computation of
transport coefficients \cite{Haas:2013hpa,Christiansen:2014ypa} is extended to the
computation of bulk viscosity and relaxation time. Moreover, the
systematic error is reduced due to a refined computation of the gluon and quark
spectral functions \cite{Cyrol:2018xeq}. Secondly, the transport coefficients are computed via
the Kubo formula from lattice results using a novel lattice approach
to imaginary time correlation functions \cite{Pawlowski:2016eck}. 
Thirdly, real-time correlation functions and transport coefficients in QCD are directly computed with non-perturbative functional methods.  For an application to QCD, real-time functional methods, see e.g. \cite{Floerchinger:2011sc,Tripolt:2013haa,Tripolt:2014wra}, have to be extended for a full numerical applications. Such an approach has been set-up in \cite{Pawlowski:2015mia,Strodthoff:2016pxx,Pawlowski:2017gxj}, and is currently applied to QCD spectral functions.
This threefold approach
is essential for increasing the reliability in the combined results by
reducing the combined systematic uncertainty, such as the absolute
normalization of the transport coefficients from the diagrammatic approach used in the present work. 
The approach to direct real-time correlations also allows for the additional treatment of the non-equilibrium corrections to the thermal distribution functions. If known from first QCD principles, this largely reduces the systematic uncertainty of the hydrodynamic description.

\section*{Acknowledgement}
The authors thank Raju Venugopalan for helpful discussions.
This work is part of and supported by the DFG Collaborative Research Centre "SFB
1225 (ISOQUANT)".
BPS and CS are supported under DOE Contract No. DE-SC0012704. 
Computational resources have been provided by the GSI Helmholtzzentrum f{\"u}r  Schwerionenforschung.



%\section*{References}

 \bibliographystyle{elsarticle-num} 
\bibliography{bib_hydro}


%\begin{thebibliography}{00}


%\bibitem{}

%\end{thebibliography}

\end{document}